\documentclass{ws-procs11x85}

\usepackage{balance}

\makeindex
\begin{document}

\title{REVIEW OF SOLAR NEUTRINO EXPERIMENTS}

\author{A. BELLERIVE}

\address{Ottawa-Carleton Institute for Physics, Department of Physics,\\
Carleton University, 1125 Colonel By Drive,
Ottawa, K1S 5B6, Canada\\E-mail: alain\_bellerive@carleton.ca}


\twocolumn[\maketitle\abstract{This paper 
reviews the constraints on the solar
neutrino mixing parameters with data collected by the Homestake, 
SAGE, GALLEX, Kamiokande, SuperKamiokande, and SNO experiments.
An emphasis will be given to the global solar neutrino analyses
in terms of matter-enhanced oscillation of two active flavors.
The results to-date, including both solar model dependent and 
independent measurements, indicate that electron neutrinos
are changing to other active types on route to the Earth 
from the Sun.
The total flux of solar neutrinos is found
to be in very good agreement with solar model calculations. 
Future measurements will focus on greater accuracy for mixing parameters
and on better sensitivity to low neutrino energies.}]

\baselineskip=13.07pt
\section{Introduction}

The deficit of neutrinos detected coming from the Sun compared with our 
expectations based on laboratory measurements, 
known as the Solar Neutrino Problem, has remained one of the outstanding 
problems in basic physics for over thirty years. 
It appeared inescapable that either our understanding of the energy producing 
processes in the Sun is seriously defective, 
or neutrinos, some of the fundamental particles in the Standard Model, have 
important properties which have yet to be identified. 
It was indeed argued by some that we needed to change our ideas on how energy 
was produced in fusion reactions inside the Sun. 
Others suggested that the problem arose due to peculiar characteristics of 
neutrinos such as oscillations and matter effects. 
It is then useful to review the evolution of our understanding from the data 
collected by various solar neutrino experiments. 
For completeness, new results from the Sudbury Neutrino Observatory not 
presented at the conference are included in the discussion presented here.

\section{Solar Neutrinos}

Hans Bethe suggested that the energy in the Sun was produced by nuclear reactions 
that allow hydrogen to be transformed 
into helium\rlap{.}\,\cite{bib:bethe} This nuclear time scale 
is about $10^{10}$ years and corresponds to 
the epoch over which a star evolves
in the main sequence. The Sun evolves slowly by adjusting its temperature so that 
the average thermal energy of 
a nucleus is small compared to the Coulomb repulsion an ion feels from potential 
fusion partners. The large Coulomb 
repulsion slows the nuclear rate to an astronomically long time scale. Hence the rate 
for nuclear reactions in the solar 
interior is dominated by Coulomb barriers. Through the reactions listed in 
Table~\ref{tab:BP}, 
four protons combine to form the helium nucleus containing two protons and two 
neutrons. As the protons 
only fuse to make helium under the very high density, high temperature conditions 
present at the centre of the Sun,
it is virtually impossible to reproduce these conditions in the laboratory and hence 
study directly the nuclear 
solar fusion hypothesis. Considering also that it takes a photon about ten 
thousand years to reach the surface 
of the Sun from the core, investigation of the electromagnetic spectrum considering 
the latter time scale dilutes 
all clues about the origin and production mechanism of the photons.

\begin{table}
\begin{center}
\caption[]{Neutrino production from fusion reactions in the Sun\rlap{.}\,\cite{bib:BP} The total solar 
flux at the Earth is $6.5 \times 10^{10}$ neutrinos per cm$^2$ and per second. The majority
of the solar neutrinos come from the $pp$ chain (more than 91\%); while the $^7Be$, 
$pep$, and $^8B$ chains correspond to about 7\%, 0.2\%, and 0.008\% of the total flux, 
respectively. The $hep$ contribution is minuscule and mostly neglected.}

\begin{tabular}{|c|c|c|}\hline
Reaction                                  &  Label  & Flux (cm$^{-2\!}$ s$^{-1\!}$) \\
\hline
$p \,+\, p \,\to\, ^2H \,+\, e^+ \,+\, \nu_e$       &  $pp$   &  $5.95 \times 10^{10}$ \\
$p \,+\, e^- \,+\, p \,\to\, ^2H \,+\, \nu_e$       &  $pep$  &  $1.40 \times 10^{8}$  \\
$^3He \,+\, p \,\to\, ^4He \,+\, e^+ \,+\, \nu_e$   &  $hep$  &  $9.3 \times 10^{3}$   \\
$^7Be \,+\, e^- \,\to\, ^7Li \,+\, \nu_e$           &  $^7Be$ &  $4.77 \times 10^{9}$  \\
$^8B \,\to\, ^8Be^* \,+\, e^+ \,+\, \nu_e$          &  $^8B$  &  $5.05 \times 10^{6}$  \\
\hline
\end{tabular}

\label{tab:BP}
\end{center}
\end{table}

However, somehow in the fusion process, two of the protons have to become neutrons. The 
only reactions that allow 
this to happen are caused by weak interactions responsible for nuclear beta decay and, 
each time a neutron 
is formed, there must be an associated electron neutrino produced. Neutrinos can travel 
directly from the core of 
the Sun to the Earth in a few minutes and hence provide a direct way to study processes 
by which protons form helium
in the Sun. As early as 1949 Luis Alvarez proposed that the hypothesis of nuclear reactions 
powering the Sun 
could be tested by measuring the solar neutrino flux. 

The number of neutrinos that we should expect is equal to twice the ratio of the energy 
received at the Earth in the form 
of sunshine, to the energy released when the four protons produce helium. The resulting 
number is huge: 
almost $10^{11}$ neutrinos pass through each square centimeter on Earth every second. Even 
with such numbers the 
detection proved a formidable challenge because of the very small scattering cross section 
of neutrinos on ordinary 
matter. The attraction of the measurement is nevertheless clear!

\begin{figure}
\center
\psfig{figure=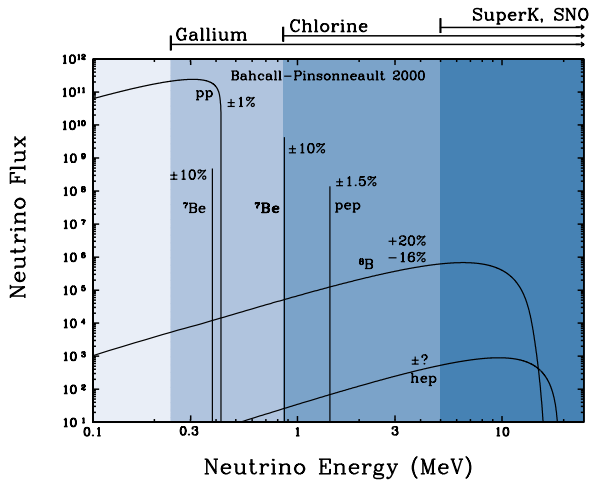,width=6.5truecm,angle=270}
\caption{The solar neutrino spectra predicted by the SSM. The neutrino 
fluxes at one astronomical unit from continuum sources are given 
in units of cm$^{-2}\!$ s$^{-1}\!$ MeV$^{-1}\!$, and the line fluxes are given 
in cm$^{-2}\!$ s$^{-1}\!$. 
Courtesy of J.N. Bahcall from http://www.sns.ias.edu/$\sim$jnb/.}
\label{fig:BP}
\end{figure}

The detailed prediction of the electron neutrino flux created by the thermonuclear reactions 
in the interior of 
the Sun was performed by John Bahcall and his collaborators from the 1960's until now. 
Their calculations are referred
to as the Standard Solar Model (SSM). In this paper, the Bahcall-Pinsonneault 
calculations\cite{bib:BP} are used
to compare experimental results and theoretical predictions. 
The solar neutrino spectra predicted by the SSM are shown in
Fig.~\ref{fig:BP}.

\section{Neutrino Oscillations}

It is known that neutrinos exist in different flavors corresponding to the three 
charged leptons: the electron, 
muon, and tau particles. If neutrinos have masses, flavor can mix in a charged-current
interaction mediated by the $W$ boson. The neutrino emitted in a weak interaction is
then a superposition of mass eigenstates 
\begin{equation}
\nu_\ell = \sum_{i=1}^n U_{\ell i} |\nu_i\rangle \, .
\end{equation}
The charged-current interactions in the leptonic sector are then described 
by the mixing matrix $U$
\begin{equation}
U = 
          \left(
          \begin{array}{cccc}
           U_{e 1}     &  U_{e 2}    & \cdots  & U_{e n}  \\
           U_{\mu 1}   &  U_{\mu 2}  & \cdots  & U_{\mu n}  \\
           U_{\tau 1}  &  U_{\tau 2} & \cdots  & U_{\tau n}
          \end{array} \right) \, .
\end{equation}
Here the neutrino mass eigenstates are denoted by $\nu_i$ with $i = 1, 2, \cdots, n$, while 
the charged lepton flavor eigenstates are labeled $(e, \mu, \tau)$. 
In the case of three generations of neutrino, the matrix $U$ is 
called the Maki-Nakagawa-Sakata-Pontecorvo (MNSP) matrix\cite{bib:mnsp} and
appears analogous to the Cabibbo-Kobayashi-Maskawa (CKM) mixing matrix\cite{bib:ckm} in the 
quark sector. The MNSP can be factorized as
\begin{equation}
U = U_{12} \times U_{23} \times U_{13} \, ,
\end{equation}
with
\begin{equation}
U_{12} =
   \left(
          \begin{array}{ccc}
              c_{12} & s_{12}  & 0  \\
             -s_{12} & c_{12}  & 0  \\
                0    &   0     & 1
          \end{array} \right) \, ,
\end{equation}
\begin{equation}
U_{23} =
   \left(
          \begin{array}{ccc}
             1    &    0    & 0      \\
             0    & c_{23}  & s_{23} \\
             0    & -s_{23} & c_{23}
          \end{array} \right) \, ,
\end{equation}
\begin{equation}
U_{13} =
   \left(
          \begin{array}{ccc}
             c_{13}  &    0    & s_{13}e^{i\delta} \\
             0       &    1    & 0                \\
-s_{13}e^{-i\delta}   &    0    & c_{13}
          \end{array} \right) \, ,
\end{equation}
where $c_{ij} = \cos\theta_{ij}$, $s_{ij} = \sin\theta_{ij}$, and $i,j$ denote
the lepton generations. Possible CP-violation is naturally embedded in the 
phase $\delta$. 

The leptonic mixing matrix naturally allows for
flavor oscillations of the neutrinos. 
The most general form for solar neutrino oscillations
can be simplified 
where only two neutrinos participate in the oscillation.
The large neutrino flavor mixing between the second
and third generation inferred from atmospheric neutrino data\cite{bib:atm} in 
conjunction with the absence of
an oscillation signal in the CHOOZ reactor neutrino experiment\cite{bib:chooz} 
requires a small component 
of one of the three mass eigenstates to the electron flavor eigenstate.
Hence the survival probability for an electron neutrino to propagate in 
time can take the approximate form
\begin{equation}
\label{eq:survive}
P_{e \alpha} = 
\delta_{e \alpha}-(2\delta_{e \alpha}-1) \sin^2 2\theta\sin^2 (1.27\frac{\Delta m^2 L}{E}) \, .
\end{equation}
The mixing angle is represented by $\theta$, $L$ is the distance between the production 
point of $\nu_e$ and the point of detection of $\nu_{\alpha}$,
$E$ is the energy of the neutrino, and $\Delta m^2 \equiv m^2_2 - m^2_1$
is the difference in the squares of the masses of the two states $\nu_2$ and $\nu_1$
which are mixing. The function $\delta_{e \alpha}$ is the usual Kronecker delta.  
The numerical constant 1.27 is valid for $L$ in meters, 
$E$ in MeV, and $\Delta m^2$ in eV$^2$. Consequently, the electron neutrino of energy $E$ 
produced in a weak interaction inside the Sun can then be described in vacuum
by the Hamiltonian
\begin{equation}
H =
   \left(
          \begin{array}{cc}
              - \frac{\Delta m^2}{4E} \cos 2\theta & \frac{\Delta m^2}{4E} \sin 2\theta  \\
                \frac{\Delta m^2}{4E} \sin 2\theta & \frac{\Delta m^2}{4E} \cos 2\theta 
          \end{array} \right) \, ,
\end{equation}
such that
\begin{equation}
i \frac{d}{dt} 
\left( 
          \begin{array}{c}
           \nu_e \\
           \nu_{\beta}
          \end{array}
\right) = 
\frac{1}{2} H 
\left( 
          \begin{array}{c}
           \nu_e \\
           \nu_{\beta}
          \end{array}
\right) \, .
\label{eq:evolution}
\end{equation}
The notation $\beta = \mu\tau$ is often used to represent the other type of 
flavor present in the solar flux. In the context of the Solar
Neutrino Problem, this suggests the investigation of (i) the disappearance 
of pure electron neutrinos produced in nuclear reactions when they reach the 
Earth, or (ii) the appearance of neutrinos of another flavor in the 
solar beam. The Earth-Sun distance is set by the planetary equations of Kepler. The energy
of the neutrino depends on the type of nuclear reaction (c.f. Table~\ref{tab:BP}) 
which produced the electron neutrino. By studying the time evolution of the 
solar neutrinos, all the
physics is then embedded in the parameters $\theta$ and $\Delta m^2$.
The full parameter space is covered with $\Delta m^2 \ge 0$ 
and $0 \le \theta \le \frac{\pi}{2}$, or
$0 \le \theta \le \frac{\pi}{4}$ and either sign for $\Delta m^2$.
Note that the survival probability of Eq.~(\ref{eq:survive}) is invariant 
under the transformations $\Delta m^2 \to - \Delta m^2$ 
and $\theta \to \frac{\pi}{2}-\theta$. These transformations
redefine the mass eigenstates by $\nu_1 \leftrightarrow \nu_2$.
This situation implies that there is a two-fold discrete ambiguity in
the interpretation of $P_{e \alpha}$ in the two-neutrino
oscillation scheme: the two different sets of physical parameters
$(\Delta m^2, \theta)$ and $(\Delta m^2, \frac{\pi}{2}-\theta)$
give the same transition probability in vacuum.

The observation of vacuum oscillations does not determine whether
$\nu_1$ or $\nu_2$ is heavier. Hence,
the measurement of $P_{e \alpha}$ in vacuum
cannot differentiate whether the larger component of $\nu_e$ resides in the
heavier or in the lighter neutrino mass eigenstate.
Thus, a solution to the Solar Neutrino Problem
implies the determination of one angle $\theta \equiv \theta_{12}$, one
mass difference $\Delta m^2 \equiv \Delta m^2_{12}$, and the
sign of $\Delta m^2$. This corresponds to the extraction
of the three MNSP elements: $U_{e1}$, $U_{e2}$, and $U_{e3}$.

\section{Matter Effects}

As was realized by Mikheev and Smirnov\rlap{,}\,\cite{bib:ms} based on the formalism
of Wolfenstein\rlap{,}\,\cite{bib:wolf} the two-fold symmetry is lost when 
mixed neutrinos travel through regions of dense matter. 
In the Sun (or the Earth), neutrinos
can undergo forward scattering with the particles
in the medium. These interactions are, in general, flavor dependent.
This possible modification of the oscillation pattern in matter is referred to as
the Mikheev-Smirnov-Wolfenstein (MSW) effect. 
When neutrinos propagate through matter the
Hamiltonian must be modified to include
scattering of the electrons in the matter,
both through neutral-current interactions (which affect
all types of neutrinos equally and lead to no observable
changes in the oscillation pattern) and through
charged-current interactions, which at solar neutrino energies
affect only the electron neutrinos. 

The conditions that have to be satisfied for the MSW effect to occur involve 
the neutrino energy, $E$, and the local electron density, $N_e$. As the process 
is a resonant one, the 
effect can be strongly energy dependent. The evolution equation is still
given by Eq.~(\ref{eq:evolution}), but
\begin{equation}
H =
   \left(
          \begin{array}{cc}
              - \frac{\Delta m^2}{4E} \cos 2\theta \,+\, \sqrt{2} G_F N_e & \frac{\Delta m^2}{4E} \sin 2\theta  \\
                \frac{\Delta m^2}{4E} \sin 2\theta & \frac{\Delta m^2}{4E} \cos 2\theta 
          \end{array} \right) \, ,
\end{equation}

\vspace{3mm}

\noindent
where $G_F$ is the Fermi coupling constant, while $\theta$ and $\Delta m^2$ are
the usual mixing parameters. In the MSW oscillation scheme, 
the matter mixing angle term, $\sin^2 2\theta_m$, is related to the
vacuum mixing angle, $\sin^2 2\theta$, by:
\begin{equation}
\sin^2 2\theta_m = \frac{\sin^2 2\theta}{(w-\sin^2 2\theta)^2 + \sin^2 2\theta} \, ,
\end{equation}
with
\begin{equation}
w = -\sqrt{2} G_F N_e E / \Delta m^2 \, .
\end{equation}

Thus, neutrinos created as electron-type in the centre of the Sun could 
emerge from the solar surface as mixed electron, muon, and tau neutrinos.
The model also implies that, under certain conditions, muon or tau neutrinos 
striking the Earth could turn back into electron neutrinos leading 
to the intriguing possibility that the Sun 
might appear brighter to neutrino 
detectors at night than during the day. 

\section{Chlorine Experiment}

The exploration of solar neutrinos started in the mid-1960's with 
Ray Davis\rlap{.}\,\cite{bib:cl} It led
to the first experiment that successfully detected neutrinos coming from the Sun. The 
experiment of Davis
and his team was carried out deep underground in the Homestake mine in the US. 
The detector was based on a concept first proposed by Bruno Pontecorvo at Chalk 
River in 1946, in which 
neutrino reactions on chlorine are measured. Neutrinos striking chlorine can make an 
isotope of argon through the 
reaction
\[
\nu_e \,+\, ^{37}Cl \to e^- \,+\, ^{37}Ar \, ,
\]
with an energy threshold of 0.814~MeV.
This reaction is rare and does not happen very often. In fact, about one atom 
of argon is produced each week in a tank containing 
100,000 gallons of the dry-cleaning fluid, perchlorethylene. The challenge of 
this radiochemical experiment is to extract the 
few atoms of argon and count them by noting their decay back to chlorine which 
occurs with a half-life of 35 days. To carry out the
low atom counting, the tank of dry-cleaning fluid is left for about a month and then 
purged with helium gas to sweep out the two or three atoms of argon. These 
atoms must then be separated from the helium by freezing them in a cold trap. They 
are then transferred to a low background counter 
where any decays are recorded over a period of several months. The first results 
were announced in 1968. The measurement 
clearly showed argon atoms produced by neutrinos, but the number was only 
one quarter of the predicted production rate. 
The chlorine experiment took data until 1995 and yielded\cite{bib:cleveland}
\begin{equation}
\Phi_{\rm{Cl}} = 2.56 \pm 0.16 \pm 0.16 \mbox{~SNU} \, ,
\end{equation}
while the SSM predicted rate was
\begin{equation}
\Phi_{\rm{Cl}}({\rm{SSM}}) = 7.6 \, ^{+ 1.3} _{-1.1} \mbox{~SNU}\, .
\end{equation}

\begin{figure}
\center
\psfig{figure=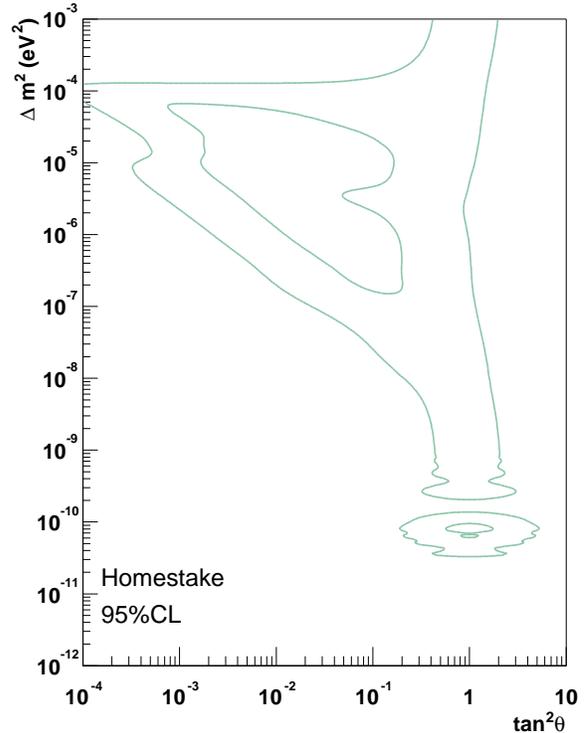,width=8.0truecm}
\caption{Mixing plane constraint provided by the chlorine experiment. The inside of the
covariance regions is allowed at the 95\% CL.}
\label{fig:cl}
\end{figure}

A SNU (Solar Neutrino Unit) is the product of the solar neutrino 
fluxes (measured or calculated) and the calculated cross sections. Hence
one SNU equals one capture per second and per $10^{36}$ target atoms.
The constraint on the oscillation parameters $\Delta m^2 - \tan^2\theta$ from 
the Homestake experiment is shown in Fig.~\ref{fig:cl}. One uses $\tan^2\theta$
instead of $\sin^22\theta$ so that $\tan^2\theta<1$ corresponds to $m_2>m_1$ (normal
hierarchy) and $\tan^2\theta>1$ to $m_2<m_1$ (inverted hierarchy). The allowed 
region is obtained by 
comparing the measured and the calculated SSM solar neutrino fluxes (see Table~\ref{tab:BP}).
Ray Davis was awarded the 2002 Nobel Prize in physics for his pioneering work which 
provided the first evidence 
that the electron neutrino flux at the Earth, created by the thermonuclear 
reactions that power the Sun, is substantially
less than would be predicted by the SSM.

\section{Gallium Experiments}

While the chlorine detector was mainly sensitive to the highest energy 
neutrinos (c.f. Fig.~\ref{fig:BP}), 
two gallium experiments, one at the Baksan laboratory\cite{bib:sage} in Russia and 
one at the Gran Sasso laboratory\cite{bib:gno} in Italy,
were set up to test the oscillation hypothesis at lower energy. 
The highest energy neutrinos and the production rate for these depends strongly 
on the solar central temperature; 
so a small change in the conditions was argued to give large changes in the 
predicted rates. On the other
hand, the lower energy neutrino flux is expected to follow directly from the 
solar luminosity as discussed above. 
The motivation of the gallium experiments was then to disentangle which MSW 
neutrino oscillation scenario
causes the Solar Neutrino Problem. 

Like the $^{37}Cl$ detector, the gallium detectors 
could only detect electron type neutrinos because they looked for the reaction
\[
\nu_e \,+\, ^{71}Ga \to e^- \,+\, ^{71}Ge \, .
\]
The energy threshold of the $^{71}Ga$ detectors is 0.233~MeV and hence allows the 
interaction of
$pp$, $^7Be$, $^8B$, and $pep$ neutrinos. The Russian-American 
group (SAGE) used a liquid metal target which 
contained 50 tons of gallium; while the European group (GALLEX/GNO) used 30 tons 
of natural gallium in an   
aqueous acid solution. Small proportional counters are used to count the germanium 
from the radiochemical target.
The $^{71}Ge$ electron capture decay occurs with a half-life of 16.5 days.
The Auger electrons and X-rays produce the typical L-peak and K-peak energy
distribution. As a cross-check, both peaks
are counted separately. A calibration with strong $^{51}Cr$ neutrino sources
provides a nice verification for low atom chemical extraction and counting 
techniques. Improvements like new electronics, better radiation shielding,
and improved calibration of counters are being pursued to continue 
regular data runs.

\begin{figure}
\center
\psfig{figure=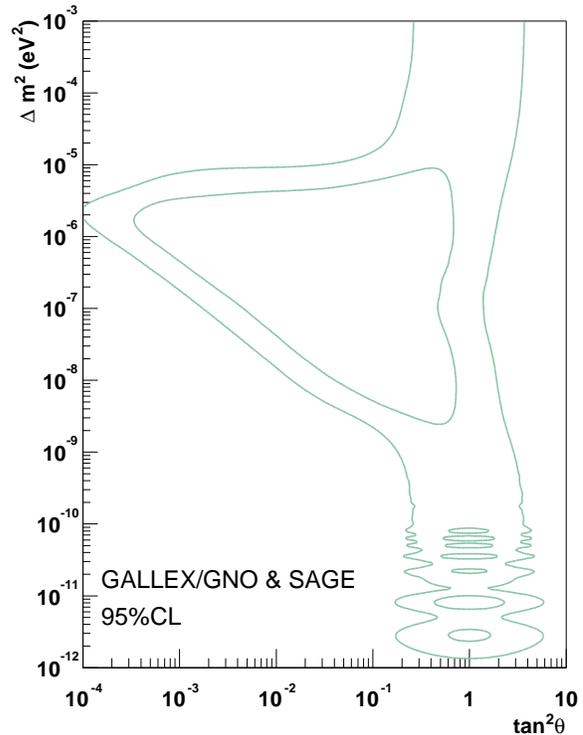,width=8.0truecm}
\caption{Constraint on the solar oscillation parameters from the gallium experiments. 
The inside of the covariance regions is allowed at the 95\% CL. This exclusion plot
uses the most recent results from SAGE and GALLEX/GNO.}
\label{fig:ga}
\end{figure}

Both experiments found about half of the expected rate. The most recent 
results of SAGE\cite{bib:lownu} were presented
at LowNu 2003 and yield for the period 1990-2003:
\begin{eqnarray}
\Phi_{\rm{Ga}} & = & 64.5 \, ^{+6.8}_{-6.5} \, ^{+3.7}_{-3.2} \mbox{~SNU~~~[L-peak]} \, ,  \nonumber \\
\Phi_{\rm{Ga}} & = & 73.3 \, ^{+5.9}_{-5.7} \, ^{+3.7}_{-3.2} \mbox{~SNU~~~[K-peak]} \, , \\
\Phi_{\rm{Ga}} & = & 69.6 \, ^{+4.4}_{-4.3} \, ^{+3.7}_{-3.2} \mbox{~SNU~~~[Overall]} \, . \nonumber
\end{eqnarray}
The GALLEX/GNO results 
for 1991-2002 were summarized at the Neutrino 2002 
conference\cite{bib:nu2002} and they were not updated for this symposium:
\begin{eqnarray}
\Phi_{\rm{Ga}} & = & 77.5 \pm 6.2 \pm 4.5 \mbox{~SNU~~~[GALLEX]} \, , \nonumber \\
\Phi_{\rm{Ga}} & = & 65.2 \pm 6.4 \pm 3.0 \mbox{~SNU~~~[GNO]} \, ,    \\
\Phi_{\rm{Ga}} & = & 70.8 \pm 4.5 \pm 3.8 \mbox{~SNU~~~[GALLEX/GNO]} \, . \nonumber
\end{eqnarray}
Yet again the data are incompatible with the SSM since 
the expected rate is
\begin{equation}
\Phi_{\rm{Ga}}({\rm{SSM}}) = 129 \, ^{+9}_{-7} \mbox{~SNU} \, .
\end{equation}
The constraint of the SAGE and GALLEX/GNO experiments on the MSW plane is 
shown in Fig.~\ref{fig:ga}. The allowed region from the combination of the 
gallium results is obtained by 
comparing the measurements with the calculations of the SSM (see Table~\ref{tab:BP}).

\section{Kamiokande and SuperKamiokande}

Following the first observations from the chlorine experiment the 
first priority
was obviously an experimental confirmation of the solar-neutrino 
deficit. This was 
provided in 1987 by the Kamiokande water \v{C}erenkov 
detector\cite{bib:kamioka} in Japan, which
also saw a significant (but, interestingly enough, not an identical) suppression 
of the measured rate of neutrinos from the Sun.
This great achievement was rewarded by the 2002 Nobel Prize in physics to 
Masatoshi Koshiba. The main advantage of the Kamiokande detector is the 
real-time nature of the neutrino
interactions viewed in the active fiducial volume (2,140 tons of ultra-pure 
light water) by 948 photomultiplier tubes (PMT). The Kamiokande Collaboration 
demonstrated that the neutrinos are
actually coming from the direction of the Sun by reconstructing the direction 
of flight of the incident neutrinos from the neutrino-electron 
scattering (ES) reaction $\nu_x \,+\, e^- \to \nu_x \,+\, e^-$. Light water detectors 
are mainly sensitive to $\nu_e$, but also to $\nu_{\mu}$ and 
$\nu_{\tau}$, with a reduced cross section
$\sigma(\nu_{\mu\tau} \, e^- \to \nu_{\mu\tau} \, e^-) \simeq 
0.15 \times \sigma(\nu_e \, e^- \to \nu_e \, e^-)$.

\begin{figure}
\center
\psfig{figure=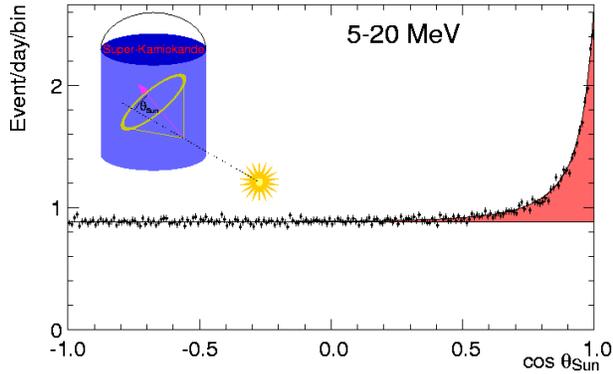,width=8.0truecm}
\caption{Angular distribution of solar neutrino candidates. SK data as of December 2002.}
\label{fig:zenith}
\end{figure}

The follow-up of the Kamiokande project is called the SuperKamiokande (SK) 
experiment\rlap{.}\,\cite{bib:SK} It was built to investigate in more detail the
nature of atmospheric and solar neutrino oscillations. 
The SK detector is a huge, 40~m in diameter and 40~m
high, circular cylinder filled with 50,000 tons of ultra-pure light water. 
The SK detector operated at an energy threshold of 5 MeV 
and hence permitted the study of the $^8B$ neutrinos.
It is divided into an outer detector to veto incoming cosmic
ray muons and to shield external low energy background; and 
an inner detector (32,000 tons, of which 22,500 tons is the active fiducial
volume) viewed by 11,146 PMT. As in Kamiokande, solar neutrinos are observed by
detecting \v{C}erenkov photons emitted by the electrons
resulting from ES events. The event rate was about 15 events per day (substantially
larger than the
rate in the radiochemical experiments).

The number of \v{C}erenkov photons collected by the PMT can be calibrated 
as a measurement of the electron energy, while the position and
times of the hit phototubes can be used to reconstruct
the $\nu_x-e^-$ interaction vertex and the electron direction. As
mentioned before and as depicted in Fig.~\ref{fig:zenith}, the electron 
direction shows a strong peak pointing directly away from the Sun
and enables discrimination against
low-energy backgrounds in the detector
arising from radioactive contaminants and
spallation products produced by penetrating cosmic
ray muons\rlap{.}\,\cite{bib:sktime,bib:1496}

\begin{figure}
\center
\psfig{figure=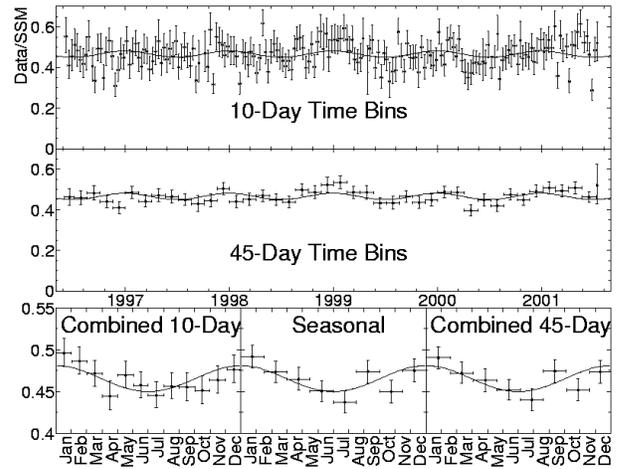,width=8.0truecm}
\caption{Time variation of the ES flux scaled by the SSM prediction. 
The curves represent the expected flux modulation due to the eccentricity of 
the Earth's orbit. SK data as of December 2002.}
\label{fig:sktime}
\end{figure}

The SK data allows measurements of the time dependence of
the ES flux. It led to the measurement of
the day/night rate asymmetry\cite{bib:1496}
\begin{equation}
A_{\rm{DN}}
= 2 \frac{\Phi_{\rm{D}}-\Phi_{\rm{N}}}{\Phi_{\rm{D}}+\Phi_{\rm{N}}} 
= -0.021 \pm 0.020 \, ^{+0.013}_{-0.012} \, ,
\end{equation}
and the precise determination of the ES neutrino flux\cite{bib:1496}
\begin{equation}
\Phi_{\rm{ES}} = (2.35 \pm 0.02 \pm 0.08) \times 10^6~{\rm{cm}}^{-2} {\rm{s^{-1}}} \, .
\label{eq:skes}
\end{equation}
The energy shape of the recoil electron agrees well, within
experimental errors, with that predicted from the neutrino
spectrum from the beta decay of $^8B$. 
The measurement of the absolute flux, however, is about 46.5\% of
that predicted by the SSM.

\begin{figure}[h]
\center
\psfig{figure=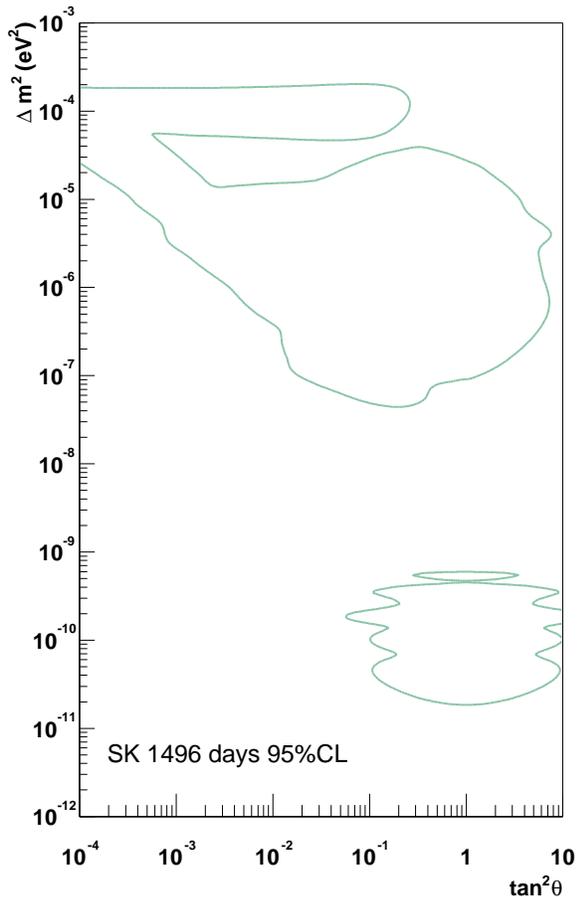,width=8.0truecm}
\caption{SSM flux independent excluded area using the SuperKamiokande
zenith spectrum shape alone. The inside of the covariance regions is excluded at
the 95\% CL.}
\label{fig:sk}
\end{figure}

The real-time data of SK also provides the framework to
study in detailed the shape of the zenith angle spectrum (i.e. the time and shape
variation of the ES energy spectrum).
The SK Collaboration looked at seasonal effects using 10 and 45 day 
bins\rlap{.}\,\cite{bib:sktime,bib:1496} This is shown in Fig.~\ref{fig:sktime}. 
The time modulation of the measured ES flux is unfortunately consistent 
with the eccentricity of the Earth's orbit around the Sun.
Therefore, beyond the rate difference between $\Phi_{\rm{ES}}$ 
of Eq.~(\ref{eq:skes}) and the SSM prediction, SK does not see any signature for
oscillation from solar neutrinos.
On the other hand, it is remarkable to visualize the complementarity of the 
radiochemical and SK experiments. Figure~\ref{fig:sk} shows the excluded regions 
of the mixing parameters from the SK Collaboration using the zenith spectrum shape
alone. The lack of spectral distortion and daily variation breaks some
degeneracy of the allowed regions of the chlorine and the gallium experiments.

Recently, SK looked for anti-neutrinos via the 
reaction $\bar{\nu}_e \,+\, p \to n \,+\, e^+$
after removing solar neutrino candidates with the effective cuts
$E>8$~MeV and $\cos\theta_{\rm{sun}}<0.5$.
From an accurate estimation of the intrinsic spallation background, which 
remains after the $\bar{\nu}_e$ selection criteria, they performed
the first search for low energy $\bar{\nu}_e$ from the Sun (which is very 
relevant if neutrinos have a magnetic moment). In the absence of a signal,
they reported an upper limit for the
conversion probability to $\bar{\nu}_e$ of the $^8B$ solar neutrinos. This 
conversion limit is 0.8\% (90\% {\rm{CL}}) of the SSM neutrino flux in the 
range of 8-20~MeV\rlap{.}\,\cite{bib:antinu}

\section{Sudbury Neutrino Observatory}

The Sudbury Neutrino Observatory (SNO) is a 1,000 ton heavy-water
\v{C}erenkov detector\cite{bib:snonim} situated 2~km underground in INCO's 
Creighton mine in Canada. Another 7,000 tons of ultra-pure light water 
is used for support and shielding. The heavy water is in 
an acrylic vessel (12~m diameter and 5~cm thick) viewed by 9,456 PMT
mounted on a geodesic structure 18~m in diameter; all contained within a 
polyurethane-coated
barrel-shaped cavity (22~m diameter by 34~m high).
The SNO detector has been filled with water since May 1999 and is moving toward 
the Neutral-Current Detector (NCD) phase of its scientific program. 
The solar-neutrino detectors in operation prior to SNO were mainly sensitive 
to the electron neutrino type; while the use of heavy water by SNO allows the flux of 
all three neutrino types to be measured. Electron neutrinos can interact through 
a charged-current interaction 
while all neutrinos can interact through a neutral-current reaction. The 
determination of these reaction rates 
is a critical measurement in determining if neutrinos oscillate in transit 
between the core of the Sun and their observation on Earth. 

Neutrinos from $^8B$ decay in the Sun
are observed in SNO from \v{C}erenkov processes following these reactions:
\begin{itemize}
\item Charged-current (CC) reaction, specific
to electron neutrinos:
\[
d \,+\, \nu_e \to p \,+\, p \,+\, e^- \, .
\]
This reaction has a Q value of 1.4 MeV and
the electron energy is strongly correlated with the
neutrino energy, providing potential sensitivity
to spectral distortions.
\item Neutral-current (NC) reaction, equally sensitive to all non-sterile 
neutrino types ($x=e, \mu, \tau$):
\[
\nu_x \,+\, d \to n \,+\, p \,+\, \nu_x \, .
\]
This reaction has a threshold of 2.2 MeV and
is observed through the detection of neutrons
by three different techniques in separate phases of
the experiment.
\item Elastic-scattering (ES) reaction:
\[
\nu_x \,+\, e^- \to \nu_x \,+\, e^- \, .
\]
This reaction has a substantially lower cross section 
than the other two and as mentioned before is 
predominantly sensitive to electron neutrinos.
\end{itemize}

The relations
\[
\Phi_{\rm{CC}}= \phi_e \, ,
\]
\begin{equation}
\Phi_{\rm{ES}}= \phi_e + 0.15 \phi_{\mu\tau} \, ,
\label{fig:nuflux}
\end{equation}
\[
\Phi_{\rm{NC}}= \phi_e + \phi_{\mu\tau} \, ,
\]
give SNO the status of an appearance experiment.
The SNO experimental plan calls for three phases
of about two years each wherein different techniques 
will be employed for the detection of neutrons from 
the NC reaction. During the first
phase, with pure heavy water, neutrons were observed 
through the \v{C}erenkov light produced when
neutrons were captured on deuterium, producing
6.25 MeV gammas. In this phase, the capture
probability for such neutrons was about 25\% and
the \v{C}erenkov light is relatively close to the threshold 
of about 5~MeV for the electron energy, imposed by
radioactivity in the detector. For
the second phase, about 2~tons of $NaCl$ was
added to the heavy water and neutron detection
was enhanced through capture on $Cl$, with
about 8.6~MeV gamma energy release and about
83\% capture effciency. Here, results from the
pure $D_2O$ and salt phases will be reported.
For the third phase, the salt will be removed and an array of
$^3He$-filled proportional counters will be installed
to provide direct detection of neutrons with a capture 
efficiency of about 45\%.

\begin{figure}
\center
\psfig{figure=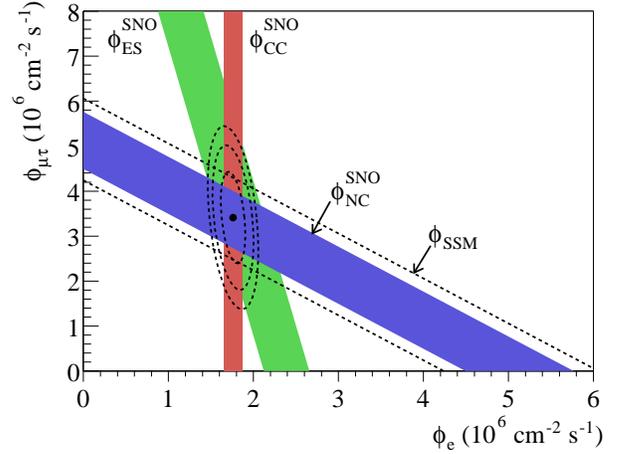,width=8.75truecm}
\caption{Flux of $^8B$ solar neutrinos which are of $\mu$ or $\tau$ flavor versus the flux 
of electron neutrinos deduced from the three neutrino reactions in SNO.  The diagonal 
bands show the total $^8B$ flux as predicted by the SSM (dashed lines) and that measured 
with the NC reaction in SNO (solid band).  The intercepts of these bands with the axes 
represent the $\pm 1\sigma$ errors.  The bands intersect at the fit values 
for $\phi_{e}$ and $\phi_{\mu\tau}$, indicating that the combined flux results are 
consistent with neutrino flavor transformation at the $5.3 \sigma$ level.}
\label{fig:snod2o}
\end{figure}

During the pure $D_2O$ phase of the experiment, the signal was extracted
with a statistical analysis technique based on the direction, $\cos\theta_{\rm{sun}}$,
the position, $R$, and the kinetic energy, $T_e$, of the event
assuming the SSM energy spectrum shape\rlap{.}\,\cite{bib:ortiz} The final 
selection criteria were $T_e \geq 5$~MeV and $R \leq 550$~cm. An
extended maximum-likelihood fit yields\cite{bib:snod2o} 
\begin{eqnarray}
\Phi_{\rm{CC}} & = & 
1.76 \, ^{+0.06} _{-0.05} \, ^{+0.09} _{-0.09} \, \times 10^6~{\rm cm}^{-2} {\rm s}^{-1} \, , \nonumber \\
\Phi_{\rm{ES}} & = & 
2.39 \, ^{+0.24} _{-0.23} \, ^{+0.12} _{-0.12} \, \times 10^6~{\rm cm}^{-2} {\rm s}^{-1} \, ,\\
\Phi_{\rm{NC}} & = & 
5.09 \, ^{+0.44} _{-0.43} \, ^{+0.46} _{-0.43} \, \times 10^6~{\rm cm}^{-2} {\rm s}^{-1} \, . \nonumber
\end{eqnarray}
The excess of the NC flux over the CC and ES fluxes implies neutrino flavor transformation.
There is also a very nice agreement between the SNO NC flux and the total $^8B$ flux 
of $5.05 ^{+1.01} _{-0.81}$ $\times 10^6~{\rm cm}^{-2} {\rm s}^{-1}$ predicted by the SSM.
The simple change of variables in Eq.~(\ref{fig:nuflux}) resolves the data directly into electron
and non-electron components\cite{bib:snod2o} 
\begin{eqnarray}
\phi_{e}       & = & 1.76 \, ^{+0.06} _{-0.05} \, ^{+0.09} _{-0.09} \, \times 10^6~{\rm cm}^{-2} {\rm s}^{-1} \, ,\\
\phi_{\mu\tau} & = & 3.41 \, ^{+0.45} _{-0.45} \, ^{+0.48} _{-0.45} \, \times 10^6~{\rm cm}^{-2} {\rm s}^{-1} \, .
\end{eqnarray}
As depicted in Fig.~\ref{fig:snod2o}, where the error ellipses 
represent the 68\%, 95\%, and 99\% joint probability contours 
for $\phi_{e}$ and $\phi_{\mu\tau}$, there is
clear evidence of solar neutrino 
flavor transformation at 5.3 standard deviations. 

\begin{figure}
\center
\psfig{figure=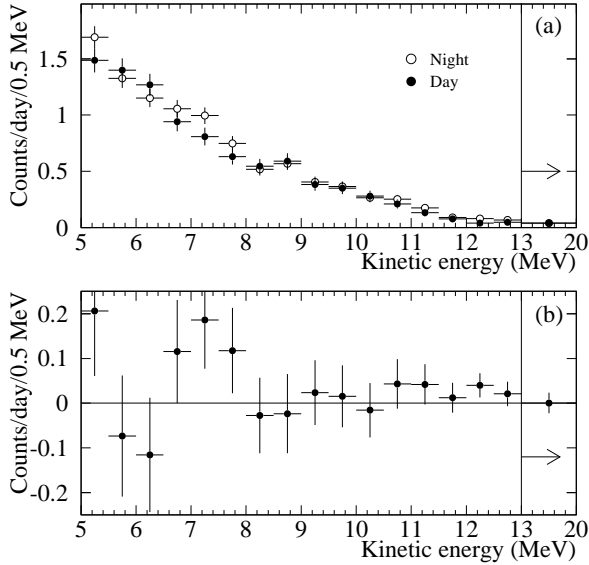,width=8.5truecm}
\caption{SNO (a) energy spectra for day and night.
All signals and backgrounds contribute. The final bin extends from
13.0 to 20.0 MeV.  (b) Difference, \rm{night - day}, between the
spectra.  The day rate was $9.23 \pm 0.27$ events/day, and the night
rate was $9.79 \pm 0.24$ events/day.}
\label{fig:snoDN}
\end{figure}

Allowing a time variation of the total flux of solar neutrino 
leads in SNO to day/night measurements which are sensitive to 
neutrino type\cite{bib:snoDN} 
\begin{eqnarray}
A_{\rm{DN}}({\rm{total}}) = (-24.2 \pm 16.1 \, ^{+2.4} _{-2.5}) \, \% \, , \\
A_{\rm{DN}}(e) = (12.8 \pm 6.2 \, ^{+1.5} _{-1.4}) \, \% \, .
\end{eqnarray}
By enforcing no asymmetry in the total rate, i.e. $A_{\rm{DN}}({\rm{total}})=0$,
the day/night asymmetry for the electron neutrino is\cite{bib:snoDN}
\begin{eqnarray}
A_{\rm{DN}}(e) = (7.0 \pm 4.9 \, ^{+1.3} _{-1.2}) \, \% \, .
\end{eqnarray}
The day and night energy spectra for all accepted events are shown in
Fig.~\ref{fig:snoDN}.
Backgrounds were subtracted separately for day and night as part of
the signal extraction. No systematic has been identified, in
either signal or background regions, that would suggest that the
small difference between day and night is other than a statistical
fluctuation.

Even if they were not ready for this conference, SNO published
their first results of the salt phase\cite{bib:snosalt} shortly 
after. The measurements were made 
with dissolved $NaCl$ in the heavy water
to enhance the sensitivity and signature for neutral-current interactions.  
Neutron capture on $^{35}Cl$ typically produces multiple $\gamma$ rays while 
the CC and ES reactions produce single electrons. The greater isotropy of
the \v{C}erenkov light from neutron capture events relative to CC and ES events 
allows better statistical separation of the event types. More importantly, this 
separation allows a precise 
measurement of the NC flux to be made independently of assumptions about 
the CC and ES energy spectra. 
The degree of the \v{C}erenkov light isotropy is determined from the pattern 
of PMT hits.  
To minimize the possibility of introducing biases, SNO performed a blind analysis 
procedure for the more model independent determination of the total 
active ($\nu_x$) $^8B$ solar neutrino flux.
The salt analysis was performed on the new data set, statistically separating 
events into CC, NC, ES, and external-source neutrons using an extended maximum-likelihood 
fit based on the distributions of isotropy, cosine of the event direction 
relative to the vector from the Sun, $\cos\theta_{\rm{sun}}$, and radius, R, 
within the detector. This analysis differs 
from the previous analyses of the pure $D_2O$ data\cite{bib:snod2o,bib:snoDN} since 
the spectral distributions of the ES and CC events are not constrained to the $^8B$ shape, but 
are extracted from the data. \v{C}erenkov event backgrounds from $\beta-\gamma$ decays 
were reduced with an effective electron kinetic energy 
threshold $T_e$ $\geq$ 5.5~MeV and a fiducial volume
with radius $R \leq 550$ cm. The extended maximum-likelihood analysis 
gives the following $^8B$ fluxes\cite{bib:snosalt}
\begin{eqnarray}
\Phi_{\rm{CC}} & = & 
1.59 \, ^{+0.08}_{-0.07} \, ^{+0.06}_{-0.08} \,\times 10^6~{\rm cm}^{-2} {\rm s}^{-1} \, , \nonumber \\
\Phi_{\rm{ES}} & = & 
2.21 \, ^{+0.31}_{-0.26} \pm{0.10} \, \times 10^6~{\rm cm}^{-2} {\rm s}^{-1} \, ,\\
\Phi_{\rm{NC}} & = & 
5.21 \pm 0.27 \pm 0.38 \, \times 10^6~{\rm cm}^{-2} {\rm s}^{-1} \, . \nonumber
\end{eqnarray}
These fluxes are in agreement with previous SNO measurements and the SSM.  
The ratio of the $^8B$ flux measured with the CC and NC reactions then provides
a strong signature of solar neutrino oscillations
\begin{equation}
\frac{\Phi_{\rm{CC}}}{\Phi_{\rm{NC}}}  =  0.306 \pm 0.026 \pm 0.024 \, . \\
\end{equation}

\begin{figure}
\center
\psfig{figure=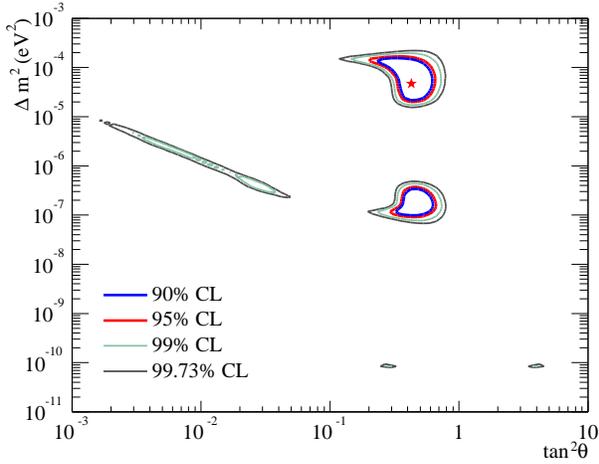,width=8.75truecm}
\caption{ SNO-only neutrino oscillation contours, including pure $D_2O$ day/night spectra, 
salt~CC, NC, ES fluxes, with $^8B$ flux free and {\emph{hep}} flux fixed.
The best-fit point is $\Delta m^{2}=4.7\times10^{-5}$, $\tan^{2}\theta=0.43$, $f_{B}=1.03$, 
with $\chi^{2}$/d.o.f.=26.2/34. The inside of the covariance regions is allowed.}
\label{fig:sno}
\end{figure}

The salt shape-unconstrained fluxes presented here combined with 
shape-constrained fluxes and day/night energy spectra from 
the pure $D_2O$ phase\cite{bib:snod2o,bib:snoDN} place 
impressive constraints on the allowed neutrino flavor mixing 
parameters. Two-flavor active neutrino
oscillation models predict the CC, NC, and ES rates in SNO.  In the fit,
the ratio $f_{B}$ of the total $^8B$ flux to the SSM value is a free parameter
together with the mixing parameters. A combined $\chi^2$ fit to SNO $D_2O$ and salt 
data alone yields the allowed regions in
$\Delta m^2$ and $\tan^2 \theta$ shown in Fig.~\ref{fig:sno}.  

\section{Global Fits}

This section summarizes the solar neutrino data in a global analysis of
all experiments. Table~\ref{tab:results} shows the experimental
results compared with the SSM prediction. The data indeed suggest an energy 
dependence due to the different energy thresholds of the different detection
techniques.

\begin{table}
\begin{center}
\caption{Overview of all the solar neutrino data. Only experimental errors 
are considered in the ratio of the data to the SSM 
predictions.}

\begin{tabular}{|c|c|c|}
\hline
Experiment           & Reaction &  Ratio data/SSM \\
\hline
Chlorine             & CC       &  0.34 $\pm$ 0.03    \\
SAGE+GALLEX/GNO      & CC       &  0.55 $\pm$ 0.03    \\
SuperKamiokande      & ES       &  0.47 $\pm$ 0.02    \\
SNO                  & CC       &  0.35 $\pm$ 0.02    \\
SNO                  & ES       &  0.47 $\pm$ 0.05    \\
SNO                  & NC       &  1.01 $\pm$ 0.13    \\
\hline
\end{tabular}

\label{tab:results}
\end{center}
\end{table}

\begin{figure}
\center
\psfig{figure=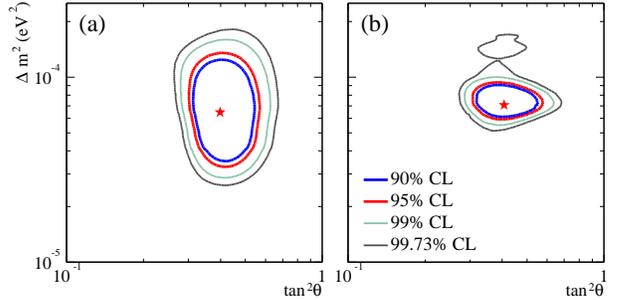,width=8.75truecm}
\caption{Allowed region of the MSW plane determined by a $\chi^2$ fit to (a) 
the chlorine, gallium, SK, and SNO experiments. The best-fit point 
is $\Delta m^2=6.5\times10^{-5}$, $\tan^{2}\theta=0.40$, $f_{B}=1.04$, 
with $\chi^{2}$/d.o.f.=70.2/81. (b) Solar global + KamLAND.
The best-fit point 
is $\Delta m^2=7.1\times10^{-5}$, $\tan^{2}\theta=0.41$, $f_{B} = 1.02$.  
In the MSW analyses, the ratio ($f_B$) of the total $^8B$ flux to the SSM value is a 
free parameter, while the total $hep$ flux is fixed to the SSM prediction.
The inside of the covariance contours is the allowed region.}
\label{fig:global}
\end{figure}

The global analysis presented here includes 
the chlorine results\rlap{,}\,\cite{bib:cleveland} the updated gallium
flux measurements\rlap{,}\,\cite{bib:lownu,bib:nu2002} the SK zenith 
spectra\rlap{,}\,\cite{bib:sktime,bib:1496} and the $D_2O$ and salt results 
from SNO\rlap{.}\,\cite{bib:snod2o,bib:snoDN,bib:snosalt}
The free parameters in the global fit are the total $^8B$ flux,
the difference of the squared mass $\Delta m^2$, and the mixing 
angle $\theta$. The higher energy $hep$ flux is fixed 
at $9.3 \times 10^3$~cm$^{-2}$~s$^{-1}$. Contours are generated in $\Delta m^2$ and $\tan^2\theta$
for $\Delta \chi^2$ = 4.61 (90\% CL), 5.99 (95\% CL), 9.21 (99\% CL), 
and 11.83 (99.73\% CL). As presented in Fig.~\ref{fig:global}(a),
the combined results of all solar neutrino experiments can be used to
determine a unique region of the MSW plane. In this 
global fit of all solar neutrino data,
the allowed regions in parameter space shrink considerably and the 
Large Mixing Angle (LMA) region is selected.

A global analysis including the KamLAND reactor anti-neutrino 
results\cite{bib:kamland} shrinks the allowed region further, with a best-fit 
point of $\Delta m^{2} = 7.1^{+1.2}_{-0.6}\times10^{-5}$~eV$^2$ and
$\theta = 32.5^{+2.4}_{-2.3}$ degrees, where the errors reflect $1\sigma$ constraints on 
the 2-dimensional region. This is summarized in Fig.~\ref{fig:global}(b).
With the new SNO measurements the allowed region is constrained to only the lower band 
of LMA at $>99\%$ CL.
The best-fit point with one dimensional projection of the uncertainties in the individual 
parameters (marginalized uncertainties) 
has $\Delta m^{2} = 7.1^{+1.0}_{-0.3}\times10^{-5}$~eV$^2$ 
and $\theta = 32.5^{+1.7}_{-1.6}$ degrees. This disfavors maximal 
mixing ($\tan^2\theta=1$)
at a level equivalent to 5.4 standard deviations.

\section{Future}

The upcoming generation of real-time low energy solar neutrino experiments will tackle 
new frontiers and allow the direct study of the mono-energetic $^7Be$ solar neutrinos.
The goal of the Borexino and KamLAND experiments is in fact to measure 
the time-dependence of the 
neutrino-electron scattering rate from neutrinos produced by electron capture on $^7Be$
in the Sun. The produced ES electrons are below threshold for production 
of \v{C}erenkov light
and thus liquid scintillator is used as detector target for enhancing the detection 
capability. The challenge at such a low
threshold is the unprecedented low level of radioactivity required; namely 
less then $10^{-16}~g/g$ 
of $^{238}U$ and $^{232}Th$ and less than one part in $10^{18}$ of $^{14}C$.
One of the dominant backgrounds to be removed is radioactive Krypton ($^{85}Kr$).
The major interest in the results for the next generation of low energy neutrino 
experiments arises when one compares the total $^7Be$ rate to the day/night 
rate asymmetry, $A_{\rm{DN}}$.
This will give an excellent independent discrimination between different MSW solutions.
It will therefore over-constrain the $\Delta m^2 - \tan^2\theta$ plane in a global analysis 
of all the solar neutrinos experiments.

The Borexino detector\cite{bib:borexino} is located in Hall C of the
Gran Sasso Laboratory. The time for the start of data taking 
is unclear due to an environmental problem encountered in 2002. Borexino is
a 300 ton liquid-scintillator based detector with 100 tons of
active fiducial mass in a 8.3~m diameter spherical nylon bag surrounded
by a 2.6 meter thick spherical shell filled with buffer oil.
The liquid scintillator and buffer liquid are viewed
by 2,240 PMT which are mounted inside a
13.5~m diameter stainless steel tank;
which is in turn surrounded by a 18~m spherical tank filled with ultra-pure
light water to act as a radiation shield. 
The expected energy threshold is about 800 keV.

The KamLAND detector is located in the cavity used for the original
Kamiokande experiment.
The primary goal of KamLAND\cite{bib:kamland} is to investigate
the oscillation of $\bar{\nu}_e$ emitted from distant nuclear 
reactors\rlap{.}\,\cite{bib:this} The investigation
of $^7Be$ neutrinos comes for free in a later stage when their scientific program will 
shift from a coincidence experiment to an ES low energy experiment. The radioactive
background remains of course the main concern for solar neutrino physics.
KamLAND hosts 1,000 tons of liquid scintillator contained in a 13~m diameter spherical balloon,
which is in turn surrounded by a 18~m spherical tank with 1,879 inward facing phototubes.
The space between the balloon and the tank is filled with buffer oil and the containment tank
is immersed in a 3,200 ton water \v{C}erenkov detector instrumented with 225 PMT.

The other important next generation of real-time
solar neutrino experiments should detect the
fundamental $pp$ neutrinos, which constitute about 91\% of the total neutrino flux predicted
by the SSM. Unfortunately, there are no approved $pp$ solar neutrino
experiments at the present time, although there are a number of promising
proposals under development and R\&D is ongoing.

\balance

\vspace{-2mm}
\section{Summary}
\vspace{-2mm}

Solar neutrino oscillation is clearly established by the combination of the 
results from the chlorine, gallium, SK, and SNO
experiments. The real-time data of SK and SNO do not show large energy distortion 
nor time-like asymmetry.
SNO provided the first direct evidence of flavor conversion of solar electron 
neutrinos by comparing the
CC and NC rates. Matter effects explain the energy dependence of solar 
oscillation, and Large Mixing Angle (LMA) solutions are favored.

The global analysis of the solar neutrino detectors and
reactor neutrino results yields 
$\Delta m^{2} = 7.1^{+1.0}_{-0.3}\times10^{-5}$~eV$^2$ and 
$\theta = 32.5^{+1.7}_{-1.6}$ degrees.  Maximal mixing is 
rejected at the equivalent 
of 5.4 standard deviations and confirms the region $\tan^2\theta<1$, which 
corresponds to the normal mass hierarchy 
of $m_2 > m_1$ (i.e. $\Delta m^2>0$).

Solar neutrino data demonstrates that neutrinos have mass and that
the minimal 
Standard Model is incomplete. Unlike the quark sector where the 
CKM mixing angles are small, the lepton sector exhibits 
large mixing. The neutrino masses and mixing may play significant roles 
in determining structure 
formation in the early universe as well as supernovae dynamics and the 
creation of matter. The coming decade will be exciting for neutrino 
physics helping to delineate the {\it{New}} Standard Model 
that will include neutrino masses and oscillations. This will lead to
precision measurements of the leptonic mixing matrix, determination of 
neutrino masses, and investigation of CP and CPT properties in
the lepton sector.
After 30 years of hard labor from the nuclear and particle physics 
community, the Solar Neutrino Problem is now becoming an industry for 
precise measurements of neutrino oscillation parameters with the next 
generation of solar neutrino and long baseline neutrino experiments.

\newpage
\section*{Acknowledgments}
This article builds upon the careful and detailed work of
many people.
Special thanks to E. Bellotti, M. Boulay, J. Formaggio, V. Gavrin, K. Graham, K. Heeger, 
R. Hemingway, A. Ianni, A. Marino, M. Nakahata, A. Poon, Y. Takeuchi,
and J. Wilkerson.
This research has been financially supported in
Canada by the Natural Sciences and Engineering Research 
Council (NSERC), the Canada Research Chair (CRC) Program,
and the Canadian Foundation for Innovation (CFI).
I am grateful to the SNO Collaboration for giving me the 
opportunity to contribute to the 2003 Lepton Photon Symposium.

\clearpage
\twocolumn[
\section*{DISCUSSION}
]

\begin{description}
\item[Luc Declais] (IPNL, France):
Can you say more about the interest of real-time low energy $pp$
neutrino experiments for solar neutrinos, because GNO will measure the total amount of neutrinos above a given threshold with an
accuracy of 7\%?
So is more or less everything is known?

\item[Alain Bellerive{\rm :}]
When I first wrote the summary for future experiments, the last point 
was ``The importance of $pp$ real-time neutrino experiments'' and I changed 
it to ``Real-time low energy neutrinos are the ultimate 
probe of the Sun and test of the Standard Solar Model''. So you're right,
at the moment if you can pin down the $^7Be$ to really high accuracy, I think 
that it will provide a really good proof of matter effects in the context
of the SSM. The $pp$ neutrinos are nevertheless an important piece of the
puzzle since the majority of the neutrinos from the Sun are produced via the
$pp$ reaction.

\end{description}

\end{document}